\documentclass[aps,preprint,showpacs,superscriptaddress,groupedaddress,nofootinbib]{revtex4}

\usepackage{hyperref}  
\usepackage{graphicx}  
\usepackage{xcolor}  
\usepackage{multirow}
\usepackage{dcolumn}   
\usepackage{bm}        
\usepackage{amssymb}   

 \newcommand{\eq}[1]{(\ref{#1})}
 
 \def\l{\left}
 \def\r{\right}
 \def\pa{\partial}
 \def\nf{n_{\!f}}
 \def\eg{e.\,g.}
 \def\ie{i.\,e.}

 \def\be{\begin{equation}}
 \def\ee{\end{equation}}
 \def\bea{\begin{eqnarray}}
 \def\eea{\end{eqnarray}}
 \def\gsim{\mathrel{\rlap{\lower0.2em\hbox{$\sim$}}\raise0.2em\hbox{$>$}}}
 \def\ksim{\mathrel{\rlap{\lower0.2em\hbox{$\sim$}}\raise0.2em\hbox{$<$}}}
 \def\bm#1{\mbox{\boldmath$#1$}}
 \def\u#1{\underline{#1}}
 \def\FigScale{1.}
\begin{document}

  \title{Tracing the pressure of the gluon plasma}
  \author{G.~Jackson}
  \email{jckgre003@myuct.ac.za}
  \author{A.~Peshier}
  \email{andre.peshier@uct.ac.za}
  \affiliation{Department of Physics, University of Cape Town, Rondebosch 7700, South Africa}

\begin{abstract}
	  Being interested in how a strongly coupled system approaches asymptotic freedom, we 
	  re-examine existing precision lattice QCD results for thermodynamic properties of the 
	  gluon plasma in a large temperature range.
	  We discuss and thoroughly test the applicability of perturbative results, on which 
	  grounds we then infer that the pressure and other bulk properties approach the free 
	  limit somewhat slower than previously thought. We also revise the value of the first 
	  non-perturbative coefficient in the weak-coupling expansion.
\end{abstract}


\maketitle

\section{Introduction \label{sec: Intro}}

Among the bulk properties of a many-body system, the pressure $p$ is of particular importance as it is directly related to the partition function (or the generating functional in the context of quantum field theory). 
In the large-volume limit, and as a function of the temperature, $p(T)$ is a thermodynamic potential which allows one to calculate all other thermodynamic properties of interest, like the entropy density $s(T) = \pa p/\pa T$ and the energy density $e(T) = sT-p$.

Given its phenomenological relevance in heavy-ion physics, the pressure and the resulting equation of state of the quark-gluon plasma have been the subject of intense research in finite-temperature QCD, which has led to remarkable progress both analytically and numerically: The expansion in the coupling $\alpha$ has been pushed to its perturbatively accessible limit ${\cal O}(\alpha^3 \ln\alpha)$ \cite{Linde:1980ts}, while non-perturbative lattice QCD calculations have become much more precise due to the possibilities of modern hardware combined with improved actions. 

The (truncated) perturbative result for the pressure of the quark-gluon plasma with $\nf$ massless quark flavors has the generic structure
\be
	p_{(n)}
	=
	p_0 \l[ 1+ \sum_{m=2}^n C_m\, \alpha^{m/2} \r] ,
	\label{p_pert}
\ee
where $p_0 = \frac{\pi^2}{90} \l( 16+\frac{21}2 \nf \r)T^4$ is the interaction-free limit.
Although the coupling $\alpha(\mu)$ is evaluated at an auxiliary scale $\mu$, a compatible scale-dependence of the factors $C_m$ ensures $\mu$-invariance of $p_{(n)}$ up to sub-leading terms, $\pa\, p_{(n)}/\pa\mu = {\cal O}(\alpha^{n+1})$.
Due to screening effects, the expansion \eq{p_pert} contains half-integer powers and some of the `coefficients' $C_m$ depend on the coupling, viz.\ $C_m = c_m + \tilde c_m \ln\alpha$ for $m = \{ 4,6 \}$.
All $c_m$, $\tilde c_m$ with $m \le 6$ are known \cite{Kajantie:2002wa}, except for $c_6$ which is the first coefficient requiring non-perturbative techniques.

In the quenched limit of QCD ($\nf = 0$), lattice calculations have not only reached a remarkable level of accuracy, but now also cover a huge temperature range from below the transition temperature $T_c$ up to $T_{\rm max} = 10^3 T_c$. 
By fitting to their numerical results the corresponding analytic expression, the authors of \cite{Borsanyi:2012ve} could refine the first crude estimate \cite{Kajantie:2002wa} of the non-perturbative coefficient $c_6$.
They also concluded that the adjusted $p_{(6)}(T)$, topped off by the adjusted $c_6$-term, remains applicable down to $T \sim 10T_c$ -- which appears remarkable for a weak-coupling approximation.

Motivated by our interest in how a strongly coupled system like a QCD plasma near $T_c$ approaches the perturbative limit, we will here re-examine this analysis of \cite{Borsanyi:2012ve}. 
After a brief summary of the existing results, we will put forward a somewhat different approach; relating, in a complementing way, lattice and perturbative results. 
On one hand, this will shed some light on the applicability of perturbation theory. 
It will also imply a small, but systematic modification of the published lattice results for the pressure (and other thermodynamic properties), and a modification of the value for the non-perturbative coefficient $c_6$.

\section{On the state of the art \label{sec 2}}

In \cite{Borsanyi:2012ve}, the pressure was evaluated not directly, but by the so-called integral method \cite{Engels:1990},
\be
	\frac{p(T)}{T^4}
	=
	\sigma + \int_{T_0}^T \frac{dT'}{T'}\, \frac{{\cal I}(T')}{T'\hskip 0.03em^4} \, ;
	\label{int method}
\ee
actually computed in this and many other lattice simulations is the trace of the energy-momentum tensor, ${\cal I} = e-3p = T^5 \pa(p/T^4)/\pa T$, also referred to as interaction measure.
Thus, for given ${\cal I}(T)$, the pressure calculated via \eq{int method} also depends on the integration constant $\sigma = p(T_0)/T_0^4$, which thus affects the `correctness' of the results.

For $T_0$ sufficiently below the transition temperature $T_c$, it is justified to approximate $\sigma \to 0$, because the confinement degrees of freedom are massive and hence thermally suppressed. 
However, integrating then to $T > T_c$ introduces a potential uncertainty: Increasing correlation lengths in the transition region -- where ${\cal I}/T^4$ peaks and thus contributes most to the integral in \eq{int method} -- lead to sizable finite-volume effects, see Fig.~2 of \cite{Borsanyi:2012ve}. 
Although these artefacts were considered in \cite{Borsanyi:2012ve} by a scaling analysis, they seem hard to correct rigorously at the aspired level of accuracy for $p(T)$ particularly for larger $T$.

In principle, this issue could be circumvented by choosing $T_0 = \infty$; the corresponding $\sigma_{\rm SB} = 16\, \frac{\pi^2}{90}$ is the familiar Stefan-Boltzmann constant for the gluon plasma. 
Then integrating {\em down} in \eq{int method} from infinity would require an extrapolation of the lattice interaction measure ${\cal I}_{\rm latt}(T)$ beyond the maximal simulation temperature $T_{\rm max}$. 
To that end, the aforementioned perturbative fit of \cite{Borsanyi:2012ve} could be used where, more precisely, a subtracted version of ${\cal I}_{\rm latt}(T)$ and its analytic counterpart derived from $p_{(6)}(T)$ were matched for $10 T_c < T < T_{\rm max}$.
This fit already improved the previous, rough estimate $c_6^{[2]} = {\cal O}(-40)$ (the superscript gives the reference) of the non-perturbative coefficient to $c_6^{[3]} = -71.8 \pm 2.9$, where the uncertainty includes the statistical error and the sensitivities on the lattice scale and the fit interval. 

We will utilize this `integrating down' idea in our approach -- bearing in mind, however, the necessity to check thoroughly the consistency of matching weak-coupling and lattice results in a regime which {\sl a priori} may not be `perturbative'.
The double-log plots in Fig.~\ref{fig Borsanyi_fit} indeed reveal somewhat larger discrepancies of the fit \cite{Borsanyi:2012ve} than the original plots may suggest: 
\begin{figure}[hbt]
	\includegraphics[scale=\FigScale]{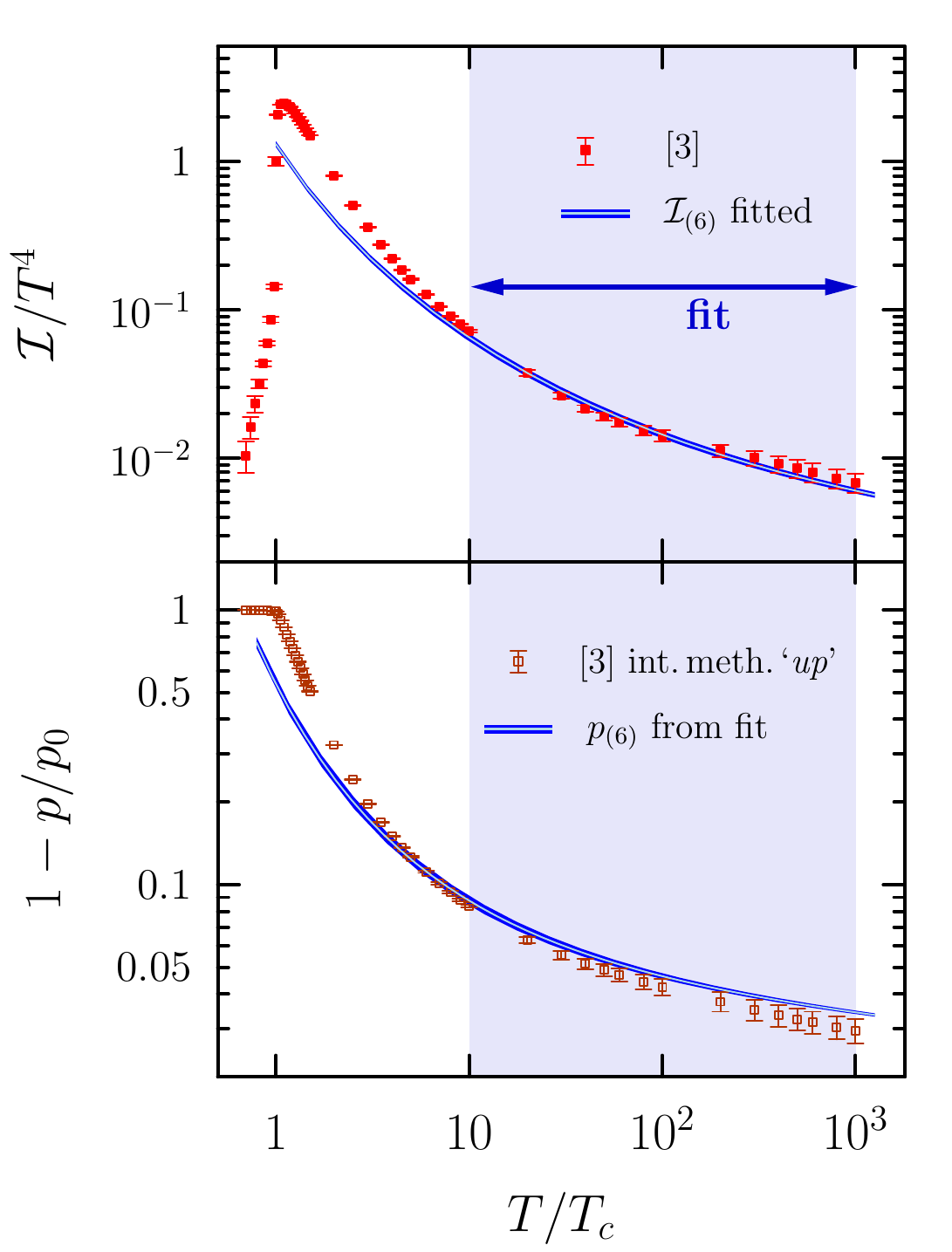}
	\caption{
	  The lattice results \cite{Borsanyi:2012ve} for the interaction measure (top), and the pressure (bottom) calculated via \eq{int method} integrating {\em `up'} as discussed in the main text. 
	  In both panels, the lines show the adjusted perturbative results obtained in \cite{Borsanyi:2012ve}, the narrow bands depict the $c_6^{[3]}$-uncertainty.
		\label{fig Borsanyi_fit}}
\end{figure}
For $T \in [10^2T_c,T_{\rm max}]$, the fit ${\cal I}_{(6)}(T)$ is systematically on the lower bound of the uncertainty band of ${\cal I}_{\rm latt}$ -- although we would expect perturbation theory to become more accurate at larger $T$. 
As a direct consequence thereof, the `lattice results' for the pressure (actually calculated from ${\cal I}_{\rm latt}$ by integrating `up' in \eq{int method}, from $T_0 < T_c$) deviate by more than their uncertainties from the adjusted $p_{(6)}(T)$ -- notably for {\em all} temperatures in the fit interval $[10T_c,T_{\rm max}]$, see Fig.~\ref{fig Borsanyi_fit} lower panel.

Such discrepancies, if they were of a similar magnitude in the physical case ($\nf = 2+1$), are probably too small to have direct phenomenological implications.
Nonetheless, it seems worthwhile to revisit not only the methodology of the integral method, but also the published results for the pressure in the quenched limit (availing from the precision of ${\cal I}_{\rm latt}$), in order to have a firm benchmark for improved analytic approaches like HTL-resummations \cite{Andersen:2014dua}.

\section{Integral method and perturbative QCD \label{sec 3}}

In order to re-examine the pressure, by integrating {\em down} in the integral method \eq{int method} from a large $T_0$, we need to address the question of how reliable perturbative results can be for a system in a regime that may or may not be `weakly coupled' -- a notion which deserves a closer look.
To avoid {\em ad hoc} assumptions on whether a given value of `the coupling' is small or large, and to extend somewhat skeptical positions (based on `non-convergence of perturbative series' arguments) that perturbation theory may not be useful at all, let us first specify some basic terminology. 
We call, in general, a certain approximate calculational scheme within a given theory a {\em model}. 
Just as the underlying theory, it contains one (or more) parameter(s) that need to be specified by measuring some observable(s) ${\cal M}$ at some value(s) $\u X$ of its independent variables. 
Thereafter, we can use our model to make predictions. 
Clearly, these predictions, either for the same observable $\cal M$ at modified $X \not= \u X$, or another observable ${\cal M}'$, should be reasonably accurate as long as the predicted quantity is `not too different' from ${\cal M}(\u X)$ -- where the model is {\em exact by definition} (irrespective of the value of `the coupling').
The primary question for phenomenology is not on convergence properties of the perturbative expansion (which we see as a sequence of models), but rather the {\em range of applicability} ${\cal R}(\u X)$ of a given model (which, for fixed parameters, depends on the observables to predict). 
How much `improved' models differ (both in parameter values and range of applicability), \ie\ the question of convergence of a perturbative series, often seems only a secondary interest.

In the context of quantum field theory, the step of fixing the parameters is called renormalization: Expressing bare parameters in the Lagrangian by renormalized ones, specified by observable(s) $\cal M$ at some renormalization scale $\u X$ (whilst handling the typical infinities from loop integrals and the occurrence of a regulator $\mu$ are merely `technical details' of renormalization). 
In our case, the model is the perturbative pressure \eq{p_pert} at a given order $n$. 
Only $n=5$ and $n=6$ (the latter depending on the value of the non-perturbative coefficient $c_6$) give monotonously decreasing functions of $\alpha$ in the relevant range of values and are thus physically meaningful.\footnote{%
	We do not consider the leading order ($n=2$) result since strictly speaking, without resummed loop insertions, the coupling should not run in this case. 
	We also mention that taking the log-enhanced terms, \eg\ $\tilde c_4 \ln\alpha \sqrt\alpha\,^4$, as a separate perturbative order would not give a monotonously decreasing pressure.
}
In addition to $p_{(n)}(\alpha)$, the scale-dependent `running' coupling needs to be specified. 
To $\ell$-loop accuracy (we will use $\ell = \{2,3\}$) a common analytic form reads \cite{Agashe:2014kda}
\be
	\alpha_{(\ell)}
	=
	\sum_{k=1}^\ell a_k(L) L^{-k} \, ,
\ee
where $L = \ln(\mu^2/\Lambda^2)$, $a_1 = 1.142$, $a_2 = -0.963\ln L$ and $a_3 = 0.414+0.812(\ln L-1)\ln L$.
For direct comparison with \cite{Borsanyi:2012ve} we also choose the auxiliary scale as $\mu = 2\pi T$, thus
\be
	L(T)
	=
	2\ln\l( \frac{2\pi}\lambda \frac{T}{T_c} \r) ,
	\label{L}
\ee
with the dimensionless model parameter $\lambda = \Lambda/T_c$ to be adjusted.

Our model is delineated by the orders $n$ and $\ell$ of the perturbative pressure and the running coupling, respectively. 
In order to fully specify it, we can fix the parameter $\lambda$ by matching either $p_{(n|\ell)} = p_{(n)}(\alpha_{(\ell)})$ or the corresponding interaction measure\footnote{%
	Note that ${\cal I}_{(n|\ell)}$ is of order ${\cal O}(\alpha^{n+1})$, as a result of differentiating the running coupling in $p_{(n)}(\alpha_{(\ell)})$.
	}
\be
	{\cal I}_{(n|\ell)} = T^5 \pa\l( p_{(n|\ell)}/T^4 \r)/\pa T
\ee
to the respective lattice result at a given `renormalization temperature' $\u T$.
Regardless of the value of the running coupling at the scale $\u T$ (which also depends on the choice of $\mu$), the applicability range ${\cal R}_{(n|\ell)}(\u T)$ of the resulting model ensues by comparing its {\em predictions} to the non-perturbative lattice results, at temperatures deviating from $\u T$.
Besides this applicability range, the {\em stability} of the model is quantified by the behavior of $\lambda$ as a function of $\u T$.

\subsection[Pressure to order \texorpdfstring{$n = 5$}{n5}]{Pressure to order $\bm{n = 5}$ \label{sec 3A}}

\begin{figure}[hbt]
	\includegraphics[scale=\FigScale]{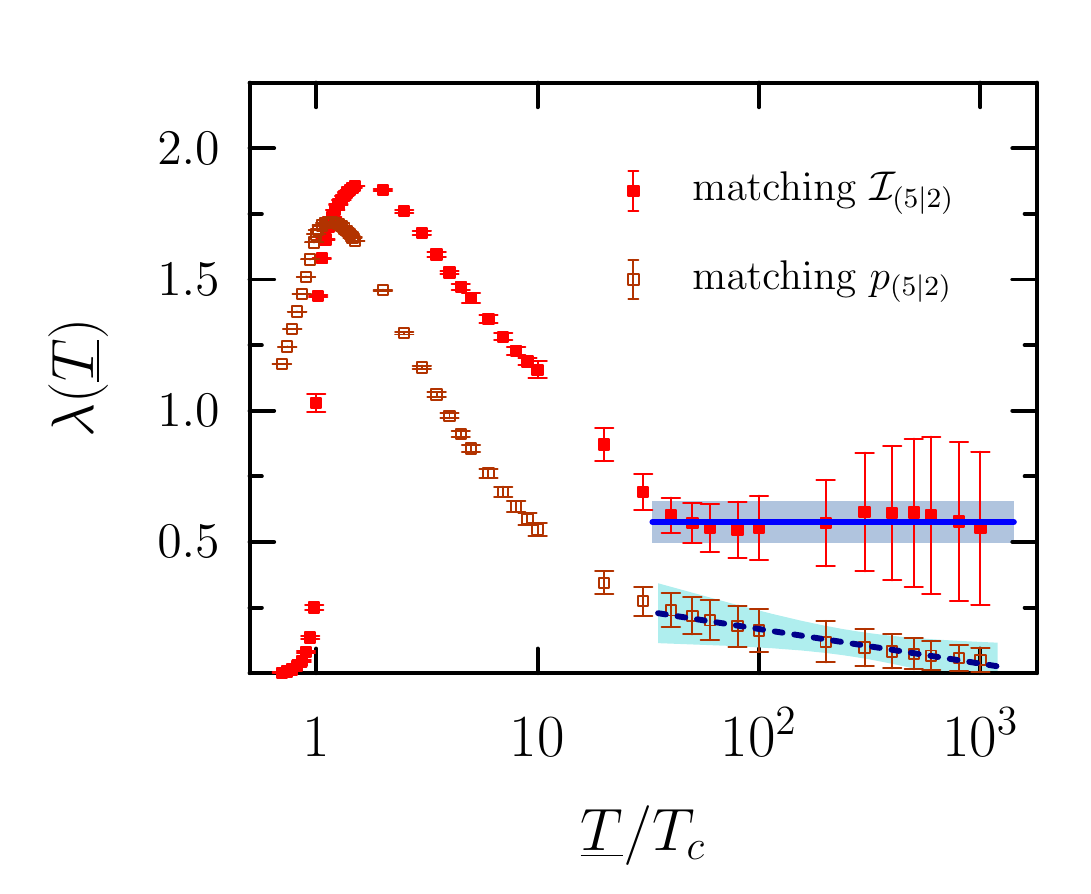}
	\caption{
	  The parameter for the $(5|2)$-model adjusted to match, at a given `renormalization temperature' $\u T$, the lattice results \cite{Borsanyi:2012ve} for either the interaction measure, cf. \eq{match I}, or the pressure, taking into account the respective uncertainties.
		\label{52_lambda}
	}
\end{figure}

\begin{figure}[hbt]
	\includegraphics[scale=\FigScale]{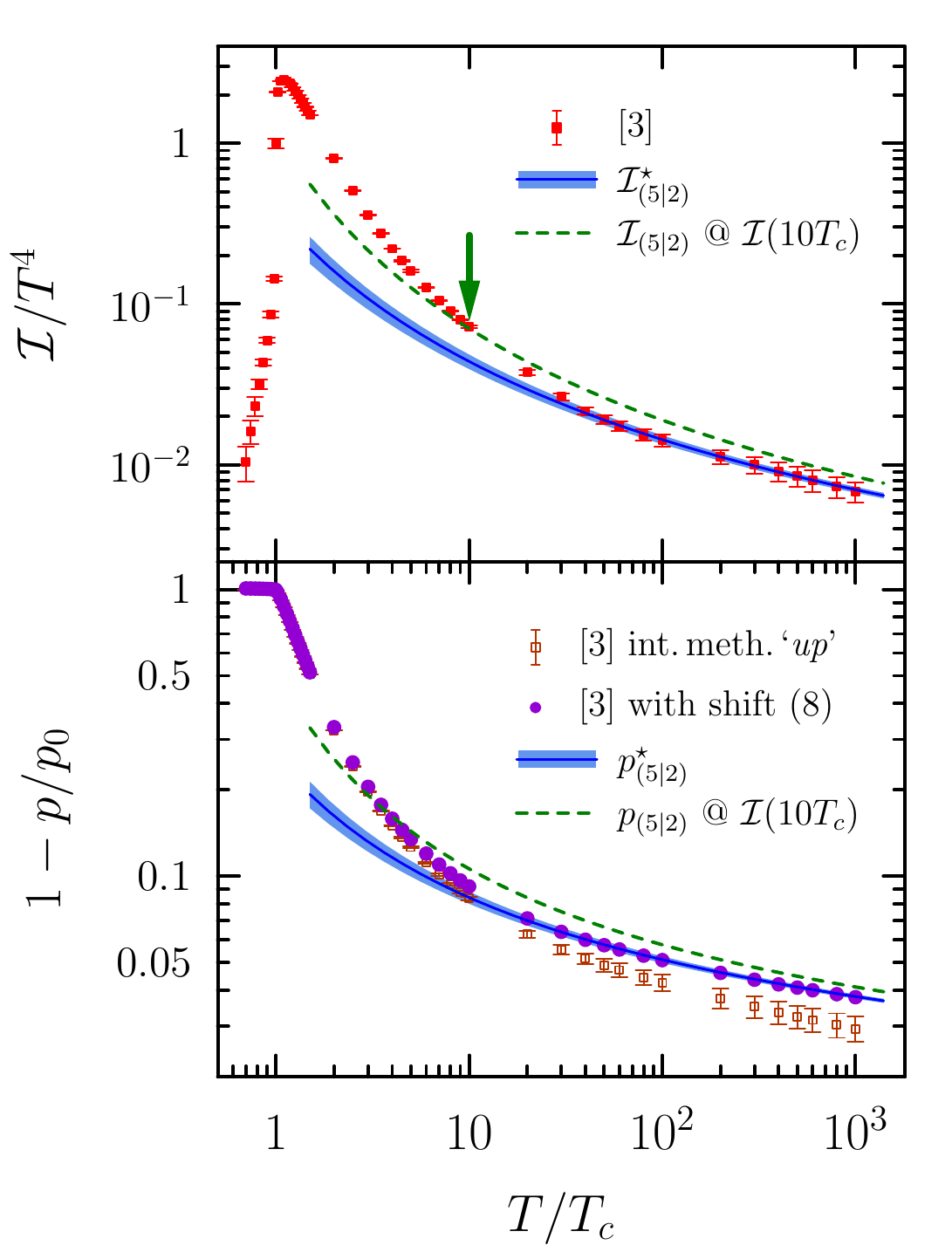}
	\caption{
	  Top: The lattice interacting measure vs.\ the $(5|2)$-model, with $i)$ adjusted parameter value \eq{lambda_52}, and $ii)$ $\lambda$ fixed to match ${\cal I}_{\rm latt}$ at $10T_c$ (indicated by the arrow) to illustrate a case out of the applicability range which, by visual inspection, is $T \gsim T^\star = 40T_c$.
	Bottom: The corresponding results for the pressure; the deviation of $p_{(5|2)}^\star(T)$ from the results obtained in \cite{Borsanyi:2012ve} is, for $T \gsim T^\star$, just a shift by the constant \eq {Delta C_52}, as depicted by the full symbols.
		\label{fig i52}
	}
\end{figure}

To order $n = 5$, all coefficients in the perturbative pressure \eq{p_pert} are known \cite{Kajantie:2002wa}.
Let us match, at a given $\u T$, the perturbative result with 2-loop coupling and the lattice interaction measure (as the primary quantity computed in \cite{Borsanyi:2012ve}), 
\be
	{\cal I}_{(5|2)}(\u T;\lambda) = {\cal I}_{\rm latt}(\u{T}) \, ,
	\label{match I}
\ee
in order to specify the sole model parameter $\lambda = \Lambda/T_c$.
Figure \ref{52_lambda} shows that the resulting $\lambda(\u T)$ is peaked with an empirical log-linear decrease for $2 \ksim \u T/T_c \ksim 30$. 
Above $T^\star = 40T_c$, $\lambda(\u T)$ is constant within the uncertainties, which underpins the perturbative stability of the $(5|2)$-model and also lets us anticipate its range of applicability. 
Combining the $\u T \ge T^\star$ set of parameter values, taking into account their uncertainties, yields
\be
	\lambda_{(5|2)}^\star = 0.58 \pm 0.11 \, .
	\label{lambda_52}
\ee
The adjusted interacting measure ${\cal I}_{(5|2)}^\star(T)$ indeed agrees with the lattice results for $T \gsim T^\star$, see Fig.~\ref{fig i52}, which confirms the anticipated applicability range ${\cal R}_{(5|2)} \approx [40T_c,\infty]$ of the model.
Figure \ref{fig i52} also illustrates the case where the $(5|2)$-approximation is inapplicable: when fixing $\lambda$ at temperatures below $T^\star$, say at $\u T = 10T_c$, the resulting range of applicability is tiny, just ${\cal O}(10\%)$ around $\u T$. 
This qualitative change in the applicability range, which mirrors the behavior of $\lambda(\u T)$ shown in Fig.~\ref{52_lambda}, lets us interpret $T^\star$ as a `cross-over' to the perturbative regime.

The $(5|2)$-model with fixed parameter $\lambda_{(5|2)}^\star$ now allows us to predict, among other observables, the pressure -- notably without the need to reconstruct it via \eq{int method}.
Given its direct relation to ${\cal I}_{(5|2)}^\star$, we expect $p_{(5|2)}^\star$ to be a reasonable approximation in a temperature range similar to ${\cal R}_{(5|2)}$, \ie\ for $T \gsim 40T_c$.
As it turns out, $p_{(5|2)}^\star(T)$ systematically deviates for {\em all} temperatures from the lattice pressure obtained in \cite{Borsanyi:2012ve}, see Fig.~\ref{fig i52} lower panel.
However, for $T > 30T_c$ -- \ie\ right in the ballpark of ${\cal R}_{(5|2)}$ -- these deviations are just a {\em constant} shift
\be
	\Delta\sigma_{(5|2)} \approx -1.5\cdot 10^{-2}
	\label{Delta C_52}
\ee	
in $p/p_0$.\footnote{%
	By contrast, for $T \ksim 30T_c$ the deviations are $T$-dependent,  indicating the breakdown of the approximation.
}
This fact lets us scrutinize, now in a more substantiated way than in Sec.~\ref{sec 2}, the `lattice pressure' as published in \cite{Borsanyi:2012ve}.
Since we have all reasons to consider the adapted perturbative result
$p_{(5|2)}^\star$ as reliable at large $T \gsim 40 T_c$, we ascribe the offset \eq{Delta C_52} to the pronounced finite-size effects of the lattice interaction measure in the vicinity of $T_c$ due to large correlation lengths, see Sec.~\ref{sec 2}.
It appears that these artefacts have been only partly corrected for in \cite{Borsanyi:2012ve}, which then leads to an accumulation of errors when integrating in \eq{int method} {\em up}, \ie\ `over the peak' of ${\cal I}_{\rm latt}/T^4$ around $T_c$.

For more accurate results for the pressure at temperatures $T \gsim 2T_c$ (where the finite-size of effects of ${\cal I}_{\rm latt}$ become negligible, see Fig.~2 of \cite{Borsanyi:2012ve}) we therefore propose: Use the integral method \eq{int method} with sufficiently large $T_0 \gg T_c$ and the integration constant $\sigma$ specified perturbatively, and then integrate {\em down} to calculate $p(T)$ also outside of the perturbative regime -- avoiding the uncertainties of ${\cal I}_{\rm latt}$ near $T_c$.
In simple words, the `lattice results' \cite{Borsanyi:2012ve} for the pressure should be corrected by the shift \eq{Delta C_52} for $T \gsim 2T_c$.

Before we are going to corroborate this idea, let us comment on the applicability range of the $(5|2)$-model. 
The lower bound $T^\star = 40T_c$ of ${\cal R}_{(5|2)}$ seems difficult to infer just from the magnitude of the coupling: $\alpha_{(2)}(40T_c) \approx 0.08$ does not differ much from, say, $\alpha_{(2)}(10T_c) \approx 0.10$ or $\alpha_{(2)}(400T_c) \approx 0.06$. 
For these coupling values the successive terms in the expansion \eq{p_pert} are of similar magnitude\footnote{%
	except for $C_4 \sqrt\alpha\,^4 \approx 0.0$ for all three temperatures considered \label{footnote C4}
}
with alternating signs, \eg\ $p_{(5|2)}^\star(40T_c) \approx p_0 [1 - 0.09 + 0.12 - 0.01 - 0.08]$, which we may tentatively interpret to sum up to approximately half of the leading-order term, $p(40T_c) \approx p_0[1-\frac12 0.09]$.
These features are indications that the $\alpha$-expansion \eq{p_pert} of the QCD pressure may have similar properties as asymptotic series -- in which case higher order terms would {\em not} improve the accuracy of the approximation, except for very small couplings, $\alpha < \alpha^\star(n)$, where the bound $\alpha^\star(n)$ {\em decreases} with the order $n$ \cite{Itzykson}.

The question arises naturally whether the change \eq{Delta C_52} of the integration constant $\sigma$ in \eq{int method}, which corresponds to an ${\cal O}(1\%)$ modification of the pressure, should be considered as relevant.
In order to answer in the affirmative, let us fix anew the scale parameter $\lambda$ -- within the {\em same} $(5|2)$-model -- by matching it instead to the `lattice pressure' as calculated in \cite{Borsanyi:2012ve} (\ie\ without the proposed amendment \eq{Delta C_52}).
In this `$p$-scheme', the resulting parameter values $\lambda(\u T)$ do not converge even at the largest matching temperatures, see Fig.~\ref{52_lambda}, opposed to becoming constant for $T \gsim 40T_c$ in the case when matching the interaction measure. 
\begin{figure}[hbt]
	\includegraphics[scale=\FigScale]{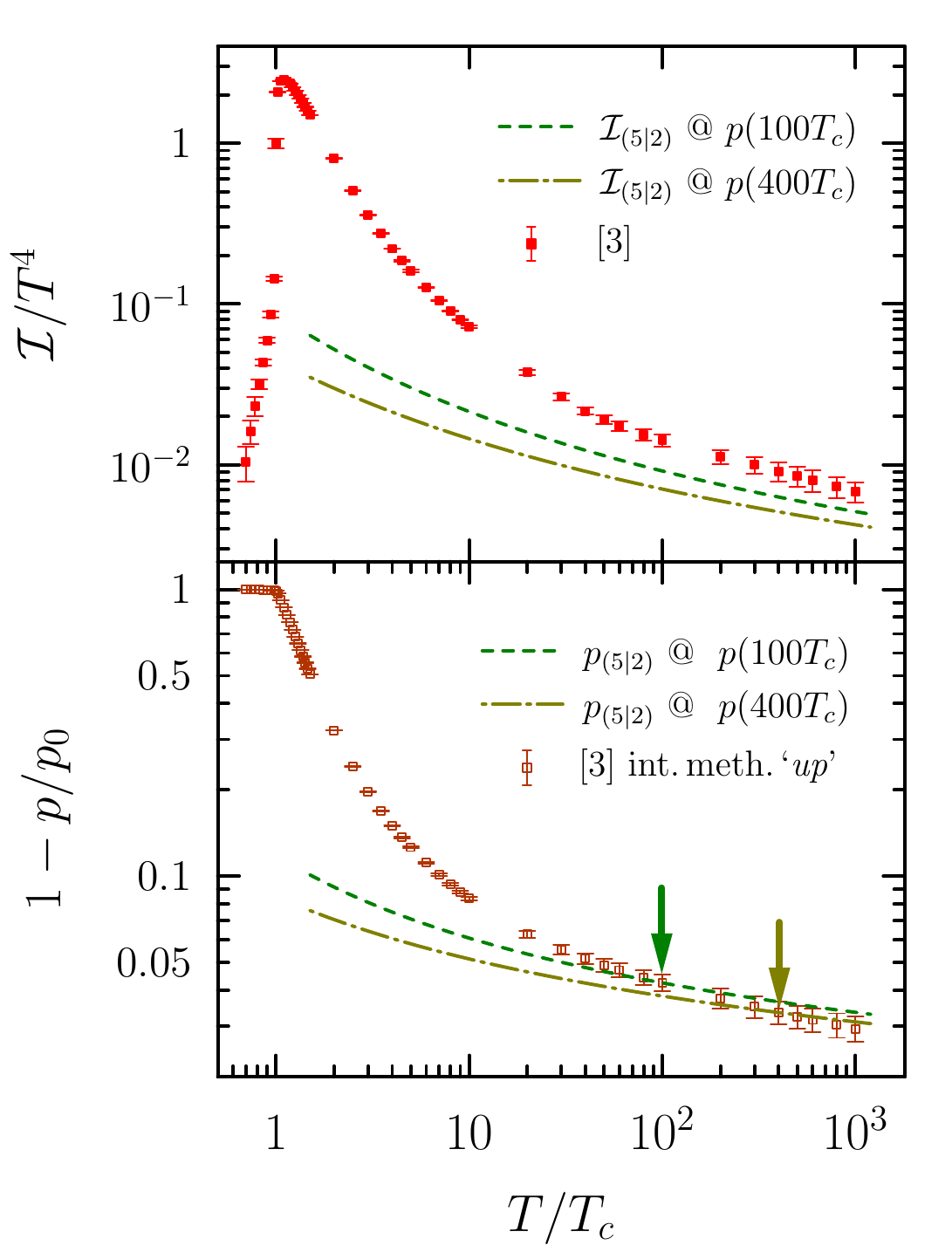}
	\caption{
	  Illustration of the $p$-scheme (to be compared to Fig.~\ref{fig i52}). 
	  Here $\lambda$ is fixed by matching $p_{(5|2)}$ to the scrutinized `lattice pressure' \cite{Borsanyi:2012ve} at $\u T/T_c = \{100,400 \}$ (indicated by arrows; bottom panel). 
	  Top: The `predicted' ${\cal I}_{(5|2)}$ clearly disagrees with ${\cal I}_{\rm latt}$ for all $T$ (even at $\u T$), notably more so with increasing `renormalization temperature' $\u T$, although one would expect perturbation theory to work better then.
		\label{fig p-scheme}
	}
\end{figure}
Lacking perturbative stability, we also expect a reduced applicability range. 
Indeed, even for the fairly large matching point $\u T = 100T_c$, we estimate ${\cal R}_{(5|2)}^p \approx [40T_c,300T_c]$ from Fig.~\ref{fig p-scheme}, lower panel, noting in particular that the upper bound clearly does not connect to the asymptotic free limit.

This striking difference between the two ways of fixing $\lambda$ is to be seen against the formal background that the magnitude of $p_{(5|2)}/T^4$ and its slope (as relevant for ${\cal I}_{(5|2)}$) are not independent; both are directly determined by $\lambda$. 
Thus, the incongruity of the $p$-scheme and the lattice results -- for {\em both} the pressure (as obtained in \cite{Borsanyi:2012ve}) and the interaction measure -- corroborates from a different perspective the importance of our proposed amendment \eq{Delta C_52} of the `lattice pressure'.

Let us emphasize that the whole picture does not change when upgrading the running coupling from 2-loop to 3-loop; in the latter case we find
\be
	\lambda_{(5|3)}^\star = 0.48 \pm 0.09 \, ,
	\label{lambda_53}
\ee
which gives virtually the same bulk properties as the $(5|2)$-model -- as expected for a stable scheme.

\subsection[Pressure to order \texorpdfstring{$n = 6$}{n6}]{Pressure to order $\bm{n = 6}$ \label{sec 3B}}
\begin{figure}[hbt]
	\includegraphics[scale=\FigScale]{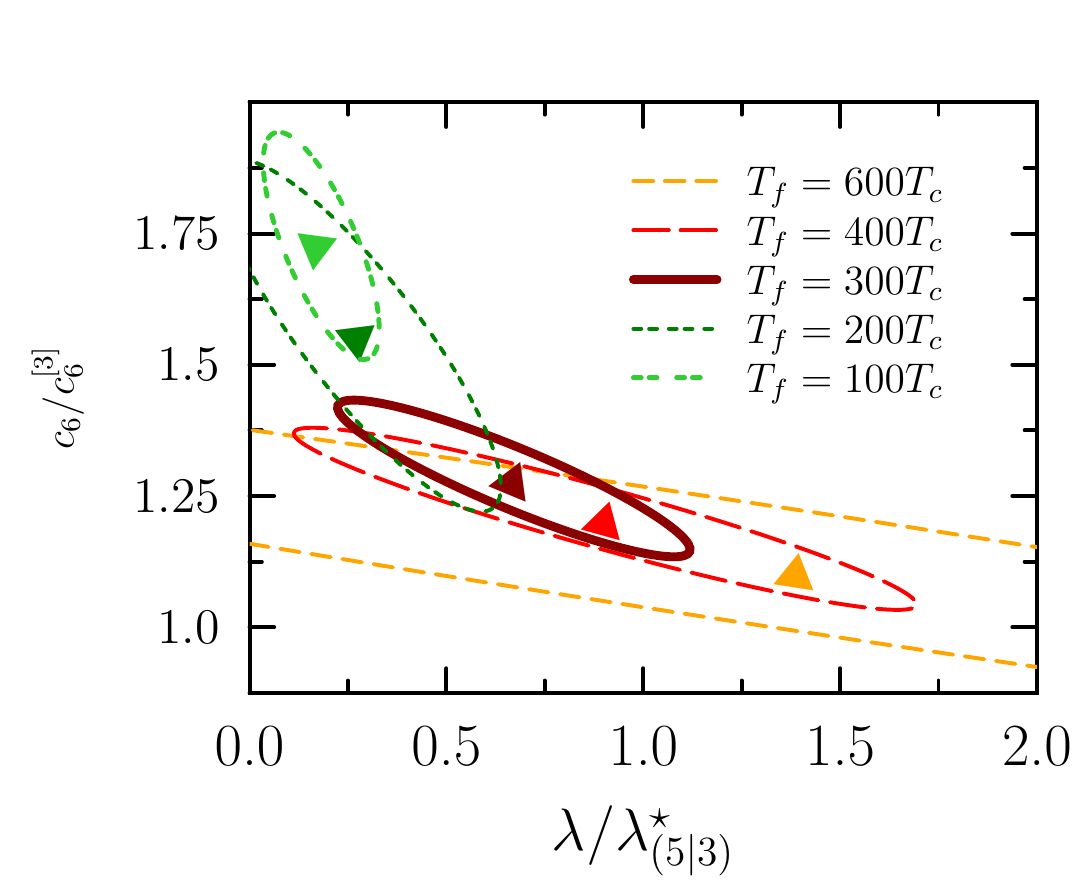}
	\caption{
	  The parameters $\{\lambda, c_6\}$ of the $(6|3)$-model with their 95\% confidence ellipses, for several fit intervals with lower bound $T_f$, and in units of $\lambda_{(5|3)}^\star$, Eqn.~\eq{lambda_53}, and $c_6^{[3]}$ from \cite{Borsanyi:2012ve}. 
	  The huge uncertainty for $T_f=600T_c$ is a consequence of having only 3 `data points' to fit.
	  For better visibility, all ellipses were stretched by a factor 10 in the direction of their short semi-axis.
		\label{fig 63-ellipses}
	}
\end{figure}

At order $n=6$, the expansion \eq{p_pert} contains the non-perturbative coefficient $c_6$ which can be seen, in the present context, as a second model parameter besides $\lambda$. 
We will specify both $\lambda$ and $c_6$ by matching ${\cal I}_{(6|3)}$ to the lattice interaction measure \cite{Borsanyi:2012ve}.\footnote{%
	The results with 2-loop running coupling are again very similar and basically amount to a 20\%-rescaling of $\lambda$, similar to \eq{lambda_52} vs.\ \eq{lambda_53}, without any visible change in the bulk properties.
}
For meaningfully small uncertainties, we will fit to several `data points' in the interval $[T_f, T_{\rm max}]$, bearing in mind (with regard to our discussion in Section \ref{sec 2}) to be cautious with presumptions on the applicability of the weak-coupling expansion at smaller $T_f$.
For $T_f \ge T_{(6)}^\star = 300T_c$, the fitted $\lambda$ is within its uncertainties consistent with \eq{lambda_53} in the $(5|3)$-model, while the non-perturbative coefficient $c_6$ (despite its strong correlation to $\lambda$, see Fig.~\ref{fig 63-ellipses}) turns out to be somewhat larger in magnitude than $c_6^{[3]} \approx -72$ found in \cite{Borsanyi:2012ve}.

For $T_f < T_{(6)}^\star$, both parameters and in particular their uncertainty characteristics shift notably, see Fig.~\ref{fig 63-ellipses}, which flags the limit of applicability of the model.
The large value of the lower limit $T_{(6)}^\star = 300T_c$ substantiates our concerns in Sec.~\ref{sec 2} on the attempt \cite{Borsanyi:2012ve} to extract $c_6$ on the presumption of the weak-coupling result to be valid down to $10T_c$.\footnote{%
	Setting $T_f = 10T_c$, we reproduce {\em formally} (since out of the validity range) the value $c_6^{[3]}$ found in \cite{Borsanyi:2012ve} by fitting a subtracted version of ${\cal I}_{\rm latt}$ in $[10T_c,T_{\rm max}]$.
}
We also note that this reduction of the applicability range, when increasing the perturbative order $n=5 \to 6$, is another argument supporting the asymptotic character of the perturbative expansion.

From the fit with $T_f = 400T_c$ (as a compromise between uncertainty and being safely in the applicability range) we estimate 
\bea
	\lambda_{(6|3)}^\star & = & 0.44 \pm 0.09 \, , \nonumber \\
	c_6^\star & = & -95 \pm 6 \, ,
	\label{lambda_63 & c6}
\eea
which is a significant change of the findings \cite{Borsanyi:2012ve}, $\lambda^{[3]} = 0.79 \pm 0.04$ and $c_6^{[3]} = -71.8 \pm 2.9$.
The differences result from the underlying fit intervals: whereas $[10T_c,10^3T_c]$ was chosen (somewhat {\em ad hoc}) in \cite{Borsanyi:2012ve}, we are lead by our analysis to a much smaller range $[400T_c,10^3T_c]$.
We note that the uncertainties of the parameters \eq{lambda_63 & c6} translate into fairly small uncertainties for the pressure, similar to our findings in Sec.~\ref{sec 3A}.
With regard to the forthcoming discussion we also point out that the adjusted contribution $(c_6^\star+\tilde{c}_6 \ln\alpha) \alpha^3$ is, in its validity range, only a tiny correction in the expansion \eq{p_pert}, in fact even smaller than the $n=4$ contribution discussed in Sec.~\ref{sec 3A}.

\section{Discussion}
Our analysis in Sec.~\ref{sec 3B} yields a revised value \eq{lambda_63 & c6} for the non-perturbative coefficient $c_6$, albeit with larger uncertainty than in the original analysis \cite{Borsanyi:2012ve}, which is a direct consequence of the small validity range ${\cal R}_{(6|3)} \approx [300T_c,\infty]$. 
One may therefore wonder if it is feasible at all to extract $c_6$ (and possibly higher order coefficients) from thermodynamic lattice calculations with sufficient precision.
In order to gain some insight into this question, let us test if we can determine numerically the {\em known} coefficient $c_4$ in \eq{p_pert}.

Fitting the two parameters of this $(4|3)$ `toy' model along the same lines as in Sec.~\ref{sec 3B} gives $\lambda_{(4|3)}^\star \approx 0.15$ consistently for all fit intervals with $T_f \ge 100T_c$, which is a disconcerting factor of three smaller than our previous results for $n=5,6$. 
This thwarts hope for perturbative stability, but more sobering seem to be the findings for $c_4$, whose fit result $0.6 \pm 0.2$ differs drastically from its actual value $16.2$ \cite{Kajantie:2002wa}.
The reason for this stark discrepancy becomes clear when recalling, from our discussion following Eq.~\eq{Delta C_52}, that the $n=4$ term $(c_4 + \tilde c_4 \ln\alpha)\alpha^2$ in the perturbative pressure is almost zero for relevant values of $\alpha$, see footnote~\ref{footnote C4}. 
Consequently, since the $n=4$ model truncates the sizable $c_5\alpha^{5/2}$ term, we now experience a large compensating effect on the model parameter $c_4$.

How does this situation change when going to order $n=5$, now with $c_5$ as a {\em third} model parameter to be `postdicted' by a fit? Then, although the fits over the few available `data points' become more challenging and we need suitable starting values for the algorithm to converge, we not only find a remarkable agreement of $\lambda$ with the value \eq{lambda_53}, but we can also reproduce within some 10\% (depending on details of the fit) the analytic values of both $c_4$ and $c_5$.
This is actually plausible by the same reasoning that just explained why $c_4$ cannot be extracted within the $(4|3)$ toy model: in the $n=5$ case, the omitted ($n=6$) contribution is sufficiently small, as noted at the end of Sec.~\ref{sec 3B}, and thus has hardly any `compensating' impact on the fit.

Against this background, let us reinspect and approve the $(6|3)$-model and its parameter values \eq{lambda_63 & c6}.
To that end, we complement the $n=6$ pressure by the next term $c_7 \alpha^{7/2}$ and adjust the extra parameter $c_7$ of the resulting $(7|3)$-model.
Despite the enlarged parameter space, the fit is virtually unchanged: $\lambda$ and $c_6$ are compatible with their $(6|3)$-values, while the $c_7 \alpha^{7/2}$ contribution is very small (zero within uncertainties) for temperatures in the applicability range ${\cal R}_{(6|3)}$. 
In other words, the $(6|3)$-model is perturbatively stable, which corroborates the parameters \eq{lambda_63 & c6}.

\begin{figure}[hbt]
	\includegraphics[scale=\FigScale]{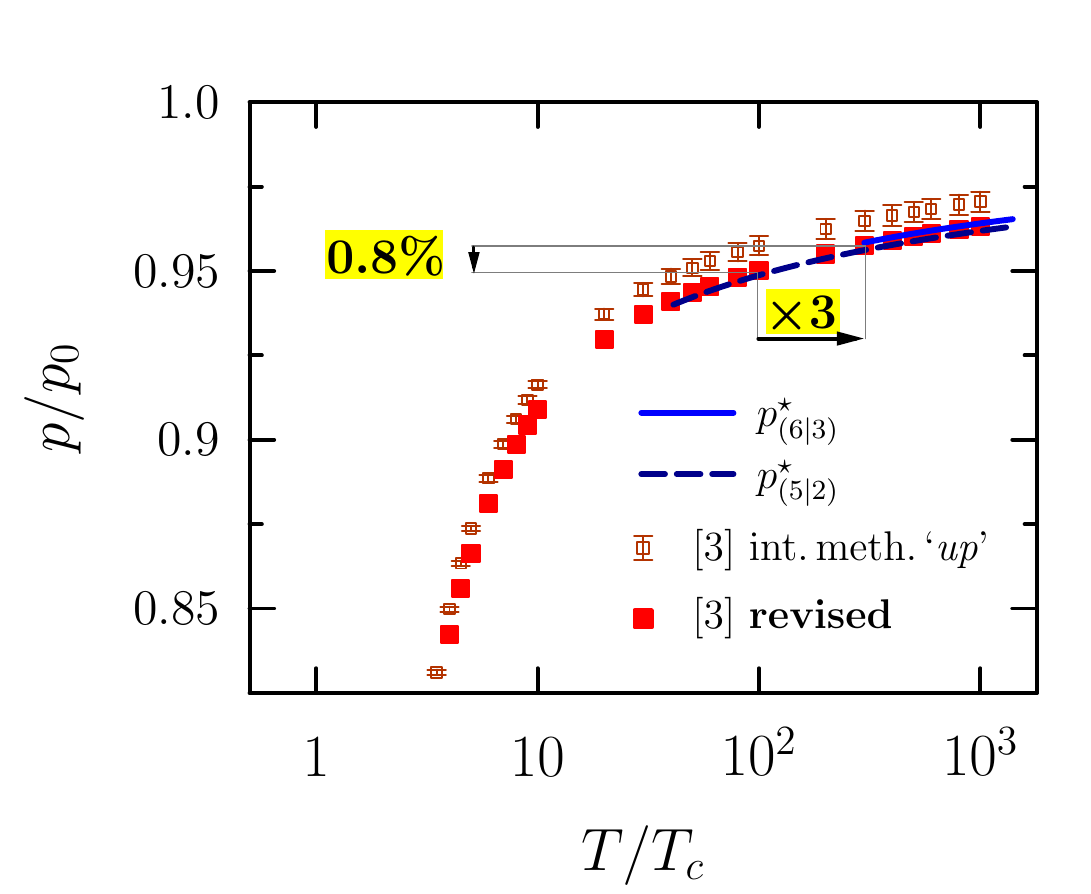}
	\caption{
	  The revised pressure calculated from the lattice interaction measure \cite{Borsanyi:2012ve}, matching the adjusted perturbative models $p_{(n|\ell)}^\star$ (results shown in the respective validity range). 
	  Note that the $0.8\%$ modification in the scaled pressure corresponds to half an order of magnitude change in the temperature scale at $T \sim 100T_c$.
		\label{fig p-amended}
	}
\end{figure}

We are now in a well-grounded position to predict the pressure in the adjusted $(6|3)$-model. 
The results $p_{(6|3)}^\star$ turn out to be very similar to $p_{(5|2)}^\star$ in the common applicability range $T \gsim T_{(6)}^\star = 300T_c$, see Fig.~\ref{fig p-amended}, in particular the almost identical {\em constant} difference $\Delta\sigma$ to the results \cite{Borsanyi:2012ve}.

After our careful analysis we therefore conclude that the results for the pressure published in \cite{Borsanyi:2012ve} need to be revised. 
Combining the modification from the $(6|3)$-model with the corresponding $(5|2)$-result \eq{Delta C_52} we find
\be
	\Delta\sigma \approx -1.4\cdot 10^{-2} \, .
	\label{Delta C}
\ee
This shift implies a $0.8\%$ reduction of the pressure at large $T$. 
In addition to the discussion at the end of Sec.~\ref{sec 3A} on the relevance of such a `small modification' in $p/p_0$, we underline here that it corresponds to a sizable factor in terms of the temperature scale, viz.\ half an order of magnitude for $T \sim 100T_c$, see Fig.~\ref{fig p-amended}.

\begin{figure}[hbt]
	\includegraphics[scale=\FigScale]{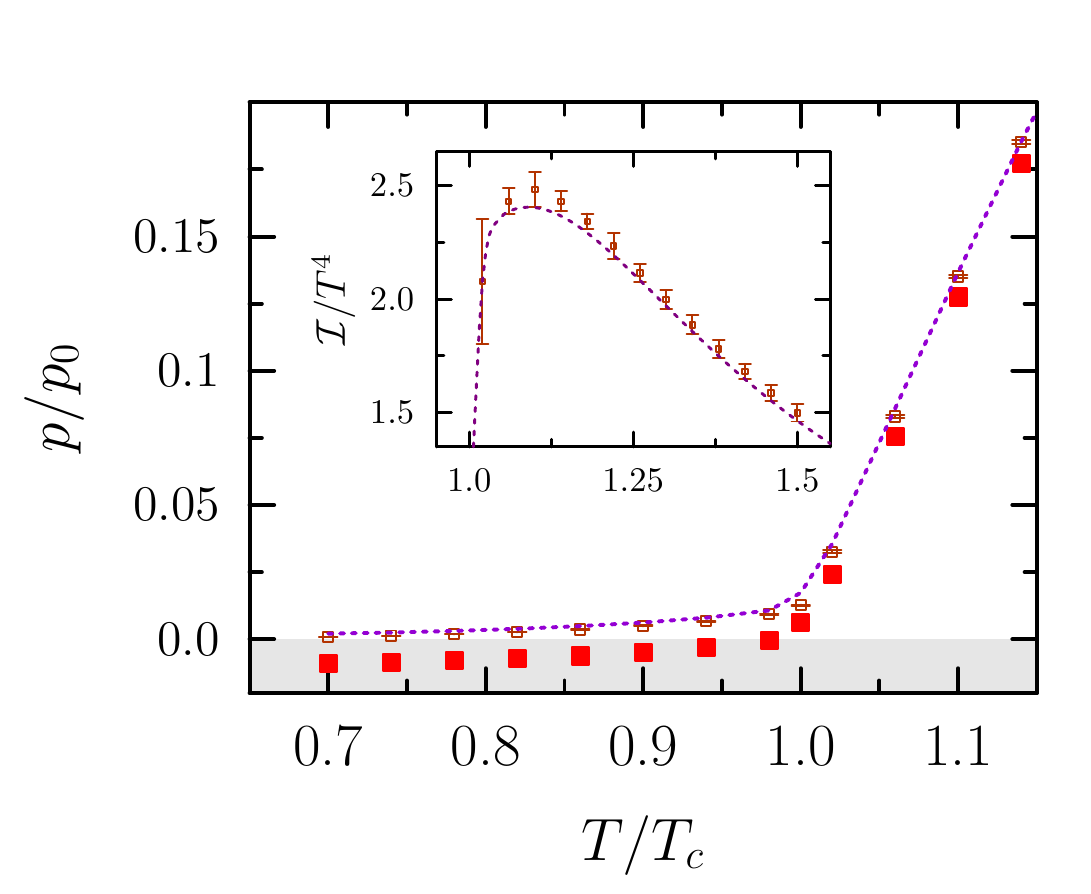}
  \caption{
    The pressure at small $T$, with the symbols matching to Fig.~\ref{fig p-amended}:
	Open squares (with almost invisibly small uncertainties) show the results \cite{Borsanyi:2012ve} from  integrating ${\cal I}_{\rm latt}$ {\em `up'}, assuming $p \to 0$ at $T \ksim 0.7T_c$. Full squares show our results from integrating  ${\cal I}_{\rm latt}$ {\em `down'}, which turn negative at $0.94T_c$.
    The expectation $p/T^4 \to 0$ can be restored by reducing the maximum of ${\cal I}_{\rm latt}$ beyond the lower end of its uncertainty (as illustrated by the inset, where error bars were scaled by 20) in the interval $T/T_c \in [1, 1.5]$; see text for details. The resulting pressure depicted by the dotted line then coincides with \cite{Borsanyi:2012ve}.}
	\label{fig 63 int down}
\end{figure}

\medskip
So far we have argued that the `lattice results' \cite{Borsanyi:2012ve} for the pressure need to be revised, based on our analysis of the interaction measure at large temperatures. 
Let us now briefly turn to necessary modifications of the interaction measure, which underlie this revision of $p$.
Relation \eq{int method}, with integration bounds $T_0 \to 0$ and $T \to \infty$ implies the `sum rule'
\be
	\int_0^\infty \frac{dT}T \frac{{\cal I}(T)}{T^4}
	\stackrel!=
	\frac{p_0}{T^4}
	=
	\sigma_{\rm SB} \approx 1.755 \, ,
	\label{sum rule}
\ee
which is a direct consequence of asymptotic freedom and confinement (the latter guaranteeing that $p(T)/T^4 \to 0$ for $T \to 0$.). 
Now, the lattice simulations \cite{Borsanyi:2012ve} cover the interval from  $T_{\rm min}=0.7T_c$ to $T_{\rm max}=10^3T_c$, which gives in \eq{sum rule} the dominant contribution 
\[
	\sigma_{\rm latt}
	=
	\int_{T_{\rm min}}^{T_{\rm max}} 
		\frac{dT}T \frac{{\cal I}_{\rm latt}(T)}{T^4}
	\approx
	1.702(2) \, ,
\]
where the error is only from numerically integrating the discrete data set (not the uncertainties) of the continuum-extrapolated lattice interaction measure.
The contribution from $T<T_{\rm min}$, which we may estimate from a simple glueball resonance gas model, is smaller than the error of $\sigma_{\rm latt}$ and can thus be omitted here. 
Consequently, the sum rule \eq{sum rule} must be saturated by the remaining large-$T$ contribution, which we have shown to be perturbative (with appropriate accuracy, for the $n=5,6$ models considered here). 
Hence
\[
	\sigma_{\rm pert}
	=
	\int_{T_{\rm max}}^\infty
		\frac{dT}T \frac{{\cal I}_{\rm pert}(T)}{T^4}
	=
	\sigma_{\rm SB} - \frac{p_{\rm pert}(T_{\rm max})}{T_{\rm max}^4} \, ,
\]
which leads to the {\em strict} constraint
\be
	\sigma_{\rm latt} < p_{\rm pert}(T_{\rm max})/T_{\rm max}^4 \, .
	\label{constraint}
\ee
It is plain from the need to recalibrate the existing lattice results for the pressure that our $(n|\ell)$-models violate this constraint (by more than the ${\cal O}(0.1\%)$ error from the numerical integration for $\sigma_{\rm latt}$), but so does even the perturbative fit obtained in \cite{Borsanyi:2012ve}. 
This discrepancy is robust: The only parameter in the $(5|\ell)$-models (which give virtually the same bulk properties as the $n=6$ models) is $\lambda$, for which we have substantiated the general expectation that it is of the order of one -- while \eq{constraint} would require $\lambda < 0.04$. 
The constraint \eq{constraint} therefore necessitates that 
the lattice results \cite{Borsanyi:2012ve} for the continuum-extrapolated interaction measure itself need to be modified. 
This revision will be `small', but necessary, as we illustrate in Fig.~\ref{fig 63 int down} from another perspective. 
Shown there is the pressure for temperatures near $T_c$, calculated by integrating ${\cal I}_{\rm latt}$ down from $T_{\rm max}$ in \eq{int method}
(with the constant
$\sigma$ fixed by our perturbative matching). 
Doing so leads to $p(T)<0$ for $T\ksim 0.94T_c$, which just reflects the inferred offset \eq{Delta C} to the findings of \cite{Borsanyi:2012ve}.
One possibility to correct this shortfall would be shifting ${\cal I}_{\rm latt}(T)$, as tabulated in \cite{Borsanyi:2012ve}, to slightly below its uncertainty band for the entire temperature range covered.
A more plausible alternative, in our view and against the background of Sec.~\ref{sec 2}, 
is that the continuum extrapolation of ${\cal I}_{\rm latt}/T^4$ is to be reduced exclusively 
near its peak, say for $[T_c, 1.5T_c]$, where finite-size artefacts are largest and thus perhaps more difficult to eliminate than by the prescription used in \cite{Borsanyi:2012ve}. 
In this case, restoring the limit $p/T^4 \to 0$ for small-$T$ and the sum rule \eq{sum rule} requires a reduction of the published continuum extrapolation of ${\cal I}_{\rm latt}$ by a few percent.

\section{Conclusions}

We put forward a systematic revision of existing lattice QCD results for the pressure of the gluon plasma, which is relevant in particular to understand the approach to the asymptotically free limit.
The scrutinized results were obtained via the integral method \eq{int method}, by integrating over a region around the confinement temperature $T_c$ where finite-size artefacts are most pronounced, which leads to an accumulation of errors at larger $T$. 
By fitting ${\cal I}_{(6)}$ to a range of values $10T_c < T < 10^3 T_c$, the corresponding pressure obtained in \cite{Borsanyi:2012ve} seems inconsistent with these errors (see Fig.~\ref{fig Borsanyi_fit}).
We rather match the lattice interaction measure (the quantity actually computed) directly to its perturbative counterpart at sufficiently large temperatures. 
The corresponding perturbative pressure determines the integration constant $\sigma$ in \eq{int method} at large $T$, which then allows us to apply the integral method {\em `down'} from the free limit (known with certainty).

\medskip
Our revision of the pressure rests on a careful analysis of the applicability of QCD perturbation theory (at the relevant orders), making use of a `thermodynamic renormalization', \ie\ fixing the QCD parameter $\Lambda = \lambda T_c$ at some temperature (range), which we then can classify as perturbative or not.
As a spinoff of the analysis, we have demonstrated that the perturbative expansion \eq{p_pert} has some characteristic features of an asymptotic series, especially for relevant orders $n$ an applicability range which {\em decreases} with increasing $n$: The order ${\cal O}(\alpha^{5/2})$ result is reliable down to some $40T_c$ vs.\ a lower limit $300T_c$ at ${\cal O}(\alpha^3)$. 
This insight has direct impact on the determination of $c_6$ as the first coefficient in the weak-coupling expansion \eq{p_pert} that is not accessible by perturbative methods. 
After the previous value was biased by including too low temperatures in the fit, we find a 30\% amendment, see \eq{lambda_63 & c6}. 
Our result seems robust from checking that the sub-leading correction $c_7\alpha^{7/2}$ is small, although the range of available large-temperature lattice results seems too narrow for a meaningful estimate of the value of $c_7$.

Presumably, our revision of the `lattice pressure' (a `small' effect and more relevant for larger temperatures, see Fig.~\ref{fig p-amended}) will not have immediate phenomenological implications for heavy-ion physics (if quenched results were directly applicable to the physical case). 
However, it seems important to be taken into account when benchmarking resummation-improved methods (see \cite{Andersen:2014dua} for a recent overview), which then in turn could provide further insight also at smaller temperatures.

The situation is different for the physical case, with $2+1$ quark flavors, where available lattice results $i)$ suffer from larger finite-size effects and $ii)$ do not yet cover a temperature range as large as in the quenched limit considered here. 
We will discuss implications of our ideas for the physical case, and phenomenological implications, in a forthcoming study \cite{Elboghdadi}.

\acknowledgments 

We thank the South African National Research Foundation and the National Institute for Theoretical Physics for support.

\end{document}